\documentclass[%
 aip,
rsi,%
 amsmath,amssymb,
 reprint,%
]{revtex4-1}
\usepackage{amssymb, amsmath, amsbsy}
\usepackage{mathdots} 
\usepackage{mathrsfs} 
\usepackage{stackrel} 
\usepackage{graphicx}
\usepackage{dcolumn}
\usepackage{bm}
\usepackage{lipsum}
\usepackage{float}
\usepackage{natbib}
\usepackage{fixltx2e}
\usepackage[utf8]{inputenc}
\usepackage[T1]{fontenc}

\usepackage{caption, subcaption}

\begin{document}

\preprint{AIP/123-QED}

\title[Homogeneity of ALD TiN MKIDS]{On the Homogeneity of TiN Kinetic Inductance Detectors \\ Produced through Atomic Layer Deposition}

\author{Israel Hernandez}
 \affiliation{Universidad de Guanajuato, Guanajuato, Mexico.}

\author{Martin Makler}%
 
\affiliation{Centro Brasileiro de Pesquisas F\'isicas, Rio de Janeiro, RJ, Brazil.}%

\author{Juan Estrada}%

\affiliation{ Fermi National Accelerator Laboratory, Batavia, IL, United States.}%



\author{Cl\'ecio R. Bom}%

\affiliation{Centro Federal de Educação Tecnol\'ogica Celso Suckowda Fonseca, Rodovia M\' arcio Covas, Lote J2, QuadraJ, Itagua\' i, Brazil.}%

\author{Donna Kubik}%

\affiliation{ Fermi National Accelerator Laboratory, Batavia, IL, United States.}%

\author{Julian Amette}%

\affiliation{Universidad de Buenos Aires.}

\author{Jorge Montes}%

\affiliation{ Fermi National Accelerator Laboratory, Batavia, IL, United States.}

\author{Andrew Lathrop}%

\affiliation{ Fermi National Accelerator Laboratory, Batavia, IL, United States.}

\begin{abstract}
The non-homogeneity in the critical temperature $T_{c}$ of an Microwave Kinetic Inductance Detector (MKID) could be caused by non-uniformity in the deposition process of the thin superconducting film. This produces low percent yield and frequency collision in the readout of the MKIDs. Here, we show the homogeneity that offers Atomic Layer Deposition (ALD). We report an improvement of up to a factor of 50  in the fractional variation of the $T_{c}$ for  TiN MKIDs fabricated with Atomic Layer Deposition in comparison with MKIDs fabricated with sputtering. We measured the critical temperature of 48 resonators.  We extracted the $T_{c}$ of the MKIDs by fitting the fractional resonance frequency to the complex conductivity of their resonators. We observed uniformity on the critical temperature for MKIDs belonging to the same fabrication process,  with a maximum change in the $T_{c}$ of 60 mK for MKIDs fabricated on different wafers.

\end{abstract}

\maketitle

\section{\label{sec:level1}INTRODUCTION}

MKIDs belongs to non-equilibrium superconducting devices, which principle of operation is based on the measurement of the electrodynamic response of the superconductor, by a change in the kinetic inductance when a photon is absorbed\cite{Non-equilibrium}. The  MKID has attractive features like high quantum efficiency, low theoretical noise limit, multiplex frequency domain, photon-number-resolving and energy resolving\cite{Guo:2017ukv}.

The MKID arrays designed for various instruments share many common structures. Usually, these structures include superconducting LC resonators (capacitor and inductor), transmission lines and antennas (See Figs.~\ref{fig:sonnetsimulatio} and \ref{fig:resonatomesurment}). The superconducting resonators (pixels) are the photon detectors, which active area consists of a meandered inductor that acts as the photosensor and an interdigitated capacitor (IDC).

The production process of an MKID consists of depositing a superconducting thin film on an insulating substrate and applying standard lithographic patterning techniques to produce a resonator structure. These simple single-layer structures permit the use of high-quality crystalline substrates and a wide variety of superconducting films,  providing an opportunity to achieve extremely low dissipation, desirable for precision instruments.

One of the most important features of the MKIDs is their capability to multiplex into large arrays. This capability enables the MKIDs to be used as powerful new astrophysical instruments for ground and space arrays besides of an extensive variety of applications. For example, one of the first MKIDs built for a camera is found in the experiment ARCONS \cite{Arcons2024}. In  SuperSpec\cite{Superscpec}, it is used as a component of a spectrograph, in the Olimpo\cite{OLIMPO_MKIDs} experiment to study the Cosmic Microwave Background (CMB) and recently, the use of  MKIDs has been  growing as an element of readout for Qubits\cite{SQUID_1,SQUID_2,SQUID_3,SQUID_4}. In addition, the MKIDs have been proposed as WIMP dark matter detectors \cite{WIMPMKIDS} and as devices for measuring the neutrino mass \cite{OLIMPO_MKIDs}.

In spite of the advantage that the MKIDs provides  to readout thousand of resonators coupled to one transmission line, in some cases, the MKIDs present problems such as non-uniformity, instability, low percent yield and frequency collisions between their resonators. These problems are usually believed to be associated with the process of deposition\cite{Szypryt:2014eua}.

In this sense, we are investigating ways to increase the homogeneity in the deposition to improve the per pixel performance of our devices through new superconducting material systems and fabrication techniques. The route that we are currently exploring attempts to create more uniform films through the use of Atomic Layer Deposition rather than the more traditional sputtering method. ALD is a thin-film technique based on the sequential use of a gas phase chemical process. Some of the advantages that offer the use of ALD are the thickness and composition control\cite{ATOMIC_HYSTORY}.  The superconducting material selected is TiN, whose $T_{c}$ is usually below 6K \cite{TiN_6K_Tc}. The critical temperature of the TiN depends on  the thickness\cite{Tc_Thicknees_1,Thicknes_2} or the N\raisebox{-3pt}{2} concentration\cite{TiN_6K_Tc,Tc_by_N2_1,Tc_by_N2_2}. Therefore, through the measurement of the critical temperature, we will assess the uniformity of ALD. 

In this paper will discuss the uniformity of ALD by measuring the critical temperature of 48 resonators coming from four identical MKIDs. All the MKIDs were fabricated with deposition of TiN through Atomic Layer Deposition under the same conditions. The  $T_{c}$ is extracted fitting the fractional resonance frequency change $\delta f_{r}(T)/f_{r} $ to some analytical expressions of the complex conductivity given by the Mattis-Bardeen theory. Finally, the reproducibility of the fabrication is investigated by measuring and comparing the resonance frequency $f_{r}$, loaded quality factor, $ T_{c}$ and kinetic inductance $L_{k}$ of each resonator, where the variations are tried to explain through simulations of some fabrication processes.

\section{Superconducting Resonators} 
\label{theory}
In a.c. current, the Cooper pairs of a superconductor are accelerated storing energy in the form of kinetic energy. The energy is stored inside the superconductor a short distance ($\lambda \approx $ 50 nm) in the form of a magnetic field. This effect is added to the surface impedance  $Z_{s}=R_{s}+ i\omega L_{s}$ of the superconducting material, where  $R_{s}$ describes a.c losses at angular frequency $\omega$ caused by electrons that are not in Cooper pairs and $L_{s}$ is the surface inductance due to the reactive energy flow between the superconductor and the electromagnetic field \cite{BASELMANS2007708,CalibrationwithNqp}. The $L_{s}$ in a $LRC$  resonant circuit is the inductive element, which is a function of  the temperature-dependent kinetic inductance $L_{k}$\cite{KineticInductance}, which is inversely proportional to the critical temperature of the superconducting material $ L_{k} \propto 1/T_{c}$. 

The resonance frequency\cite{Simon_Doyle_Thesis} of the  resonant circuit as a function of the kinetic inductance is given by 
\begin{equation}
f_{r}(L_{k})=\frac{1}{\sqrt{C_{g}(L_{g}+L_{k})}}\,, 
\label{Frsimulation}
\end{equation}
where $C_{g}$ is the geometrical capacitance and $L_{g}$ is the geometrical inductance.

The critical temperature of a resonator can be obtained  by fitting the fractional resonance frequency change $\delta f_{r}/f_{r} $ to the imaginary part of the complex conductivity of a superconductor~\cite{Aluminum}. The fractional resonance frequency change can be expressed  as  

\begin{equation}
\frac{\delta f_{r}}{f_{r}}=\frac{f_{r}(T)-f_{r}(0)}{f_{r}(0)}=-\frac{1}{2}\alpha \frac{\delta \sigma_{2}(T)}{\sigma_{2}(0)}\,,
\label{Shiftresonance}
\end{equation}
where  $\alpha$ is the ratio of the kinetic inductance and the total inductance of the resonator~\cite{GAO2006585}, and  $\sigma_{2}(T)$ is the imaginary part of the complex conductivity, which has the following analytical approximate expression according to the Mattis-Bardeen theory\cite{Aluminum,Simon_Doyle_Thesis}
 \begin{equation} 
\frac{\sigma_{2}(T)}{\sigma_{N}}\simeq\frac{\pi\Delta}{\hbar\omega}\left[1-2e^{-\frac{\Delta+\hbar\omega/2}{k_{b}T}}I_{0}\left(\frac{\hbar\omega}{2k_{b}T}\right)\right]\,.
\label{Mattis}
\end{equation}
 Here $\sigma_{N}$ is the normal conductivity, $I_{0}(x)$ is the 0-th order modified Bessel function, $\omega$ is the frequency of the a.c.  and $\Delta $ is the energy gap of the superconductor. 
 
 In TiN films it has been  shown that the complex conductivity increasingly deviates from conventional Mattis-Bardeen theory due to a strongly disorder of TiN \cite{DisorderedTiN}. This phenomenon is observed in Fig.~\ref{fig:CriticaltemperatureFit}. However, the  critical temperature of the superconductor is not affected by the disorder\cite{DIRTYSUPERCONDUCTORS} and we can consider that the $T_{c}$ is fixed in all resonators. Therefore, we expect the same value  of $\delta f_{r}/f_{r}$  as a function of the temperature in all resonators. 

 We are using microwave techniques to obtain the resonance frequency $f_{r}$ of each resonator. The complex transmission amplitude of a resonator  $S_{12}$ obtained from the network analyzer is fitted to the model of the total transmission \cite{Analysimethodgoodfunction,Besedin2018,COMPLEXSACATTERINGDATTA} of a notch type resonator coupled to the transmission line, given by 
\begin{equation}
S_{12}(f)=ae^{(-2\pi j f \tau)}\left(1-\frac{Q_{l}/|Q_{c}|e^{j\phi_{0}}}{1+2jQ_{l}\left(\frac{f-f_{r}}{f_{r}}\right)}\right)\,.
\label{GaoEquation}
\end{equation}

The coupling is via a Coplanar Waveguide (CPW) to covey microwave frequency signals. The geometrical dimension of the central strip and ground planes are shown in Fig.~\ref{fig:resonatomesurment}. The term $ 1 / | Q_ {c} | e ^ {j \phi_ {0}} $ takes impedance mismatches which are  caused by an  asymmetry in the transmission amplitude. $Q_{l}$  and $Q_{c}$ are the loaded quality factor and the coupling quality factor  of a resonator, respectively (see 
Appendix \ref{FittoS12}). 

\section{Design, fabrication and simulation}

We fabricated 24 identical MKIDs, containing 12 resonators each, in two Si wafers.  12  were fabricated in one wafer and the other 12 in another wafer. The 12 MKIDs were fabricated in the bottom half of the wafers. The upper half contains other kinds of MKIDs and stuff that were not used on this work. The production yield of MKIDs was of 100\% in both wafers. Whereby, all MKIDs could be mounted and useful.

\begin{figure}[H]
        \centering \includegraphics[width=8.5cm, height=7cm]{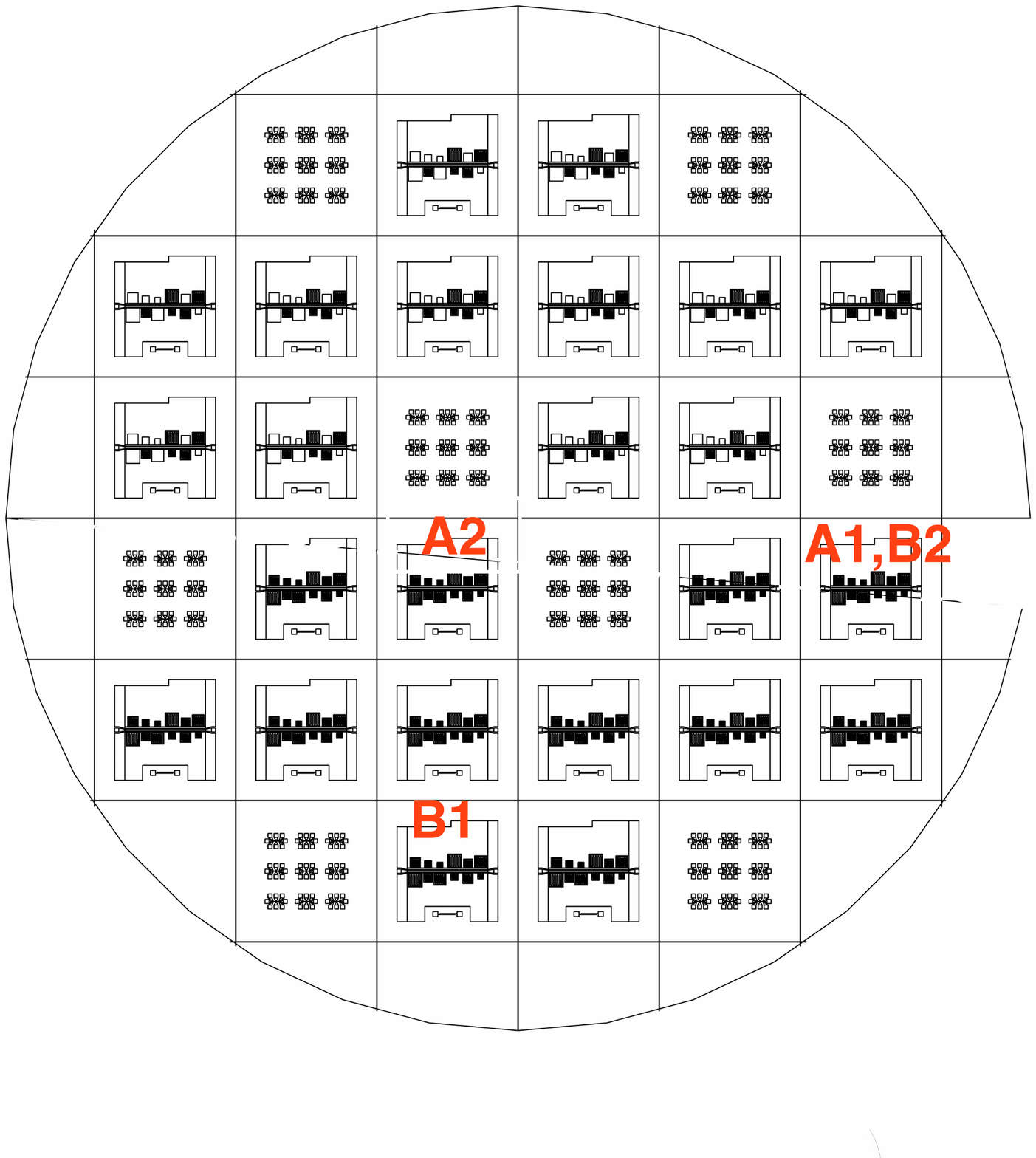}
        \caption{Positions of the MKIDs A1, B1, A2, B2 on the wafer. Wafers 1 and 2 are shown in the same image.}
       
        \label{fig:Waferposition} 
\end{figure}

We report on the comparison of four of these MKIDs, 2 from each wafer, which we will refer to A1, B1 (from wafer 1) and A2, B2 (from wafer 2).
The position of these MKIDs on the wafers is shown in Fig.~\ref{fig:Waferposition} (as they are identical, the two wafers are shown in the same figure).

\begin{figure}[H]
        \centering \includegraphics[width=\columnwidth]{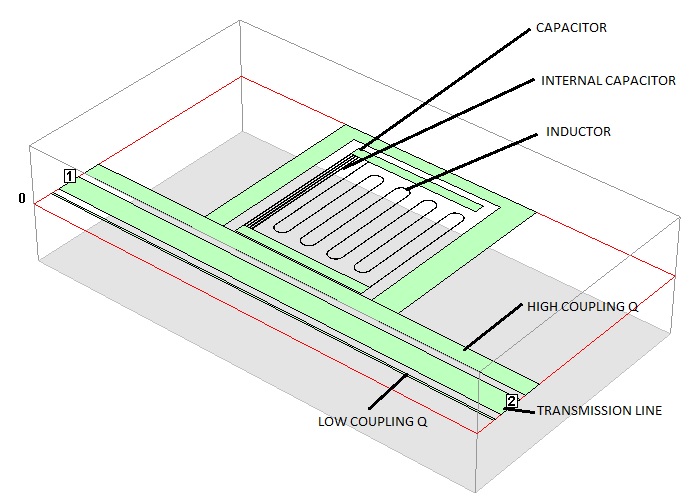}
        \caption{An example of one resonator simulated in  SONNET. There is an air layer of 1000 $\mu m$ above of the design (resonator and CPW line)  and a thickness of the Si wafer   of 500 $\mu m$ below. The thickness of the TiN is 0, value of the  surface resistance equal to 0  and variable kinetic inductance. The number 1 and 2 are the ports at 50~$\Omega$.}
\label{fig:sonnetsimulatio}
\end{figure}

\begin{figure*}[!ht]
\includegraphics[width=17.64cm, height=6.3cm]{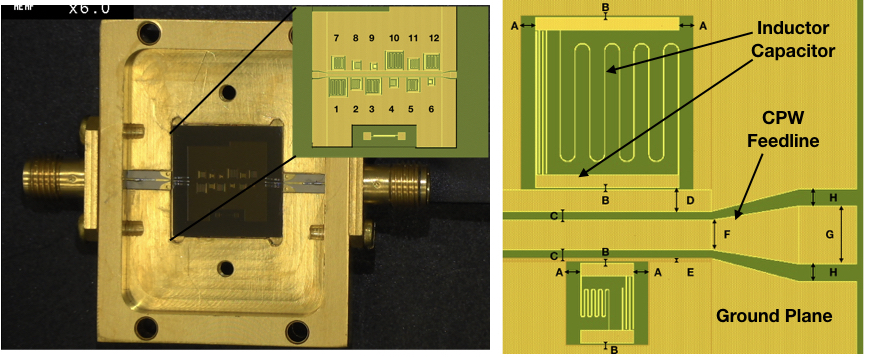}
 \caption{Left: the A1 MKID packaged and connected with SPM ports. The inset shows the position and number of each resonator. Right: zoom showing resonators 12 and 6 along with the CPW line. The dimensions A through G are stated in the text. The golden color corresponds to TiN and the green to Si.}
 \label{fig:resonatomesurment} 
\end{figure*}

The base pixel design is shown in Figs.~\ref{fig:sonnetsimulatio} and \ref{fig:resonatomesurment}. 
It consists of  one inductor in parallel with a capacitor that contains an internal interdigitated part. The system is surrounded by a ground plane that is part of a coplanar waveguide. One part of the capacitor is close to the CPW feedline and is thus coupled to the CPW.

Fig.~\ref{fig:resonatomesurment} shows a picture of an MKID in its package. The inset shows the array of 12 pixels, with different sizes. Each of the 12 resonators on an MKID was designed to have a unique resonant frequency. Six pixels (labelled 1--6) are coupled to the CPW feedline through a thick conducting stripe, providing a high quality factor $Q_{l}$.
The other six pixels (7--12) are coupled to the transmission line through a thin stripe, providing a low $Q_{r}$.

The right-hand side of  Fig.~\ref{fig:resonatomesurment} shows the details of the geometry of two resonators and the CPW. The separation with respect to the ground plane is identical for each pixel, $A=102~\mu m$ and $B=15~\mu m$ (see the labels in the figure). The CPW line is separated from the two ground planes by a distance $C=60~\mu m$. The separation of the capacitors to the CPW line varies according to the expected strength of the coupling, reflecting on the value of $Q_{l}$.
For the $Q_{r}$ resonators we have $D=150~\mu m$, whereas for the low $Q_{r}$ ones we have $E=18~\mu m$. The width of the feedline across the resonators is $F=200~\mu m$, broadening to $G=400~\mu m$ on the extremities, with a separation to the ground planes of $H=120~\mu m$.

The MKIDs fabrication was performed at University of Chicago's Pritzker Nanofabrication Facility. The fabrication process involved three main steps: {\it i}) the deposition of 300 layers of TiN on a 125~mm Si wafer (using the Ultratech/Cambridge Fiji G2 Plasma-Enhanced ALD equipment), {\it ii}) the imprint of the circuits using maskless lithography (with the Heidelberg MLAD150 Direct Write Lithographer), and {\it iii}) the removal of the excess TiN through dry etching (employing the Plasma-Therm ICP Chlorine Etch). The chosen ratio of the TiN and Si etch rates was 1:2. 
The deposited layers of TiN correspond to a thickness of 25~nm, which was chosen so that the expected superconducting transition temperature is about 4.1K.

Each of the four MKIDs was packaged in a separate gold-plated copper box (see Fig.~\ref{fig:resonatomesurment}) with 50~$\Omega$ SMA ports and each was mounted and tested under the same conditions. The copper box was mounted on the cold stage of a two-stage adiabatic demagnetization refrigerator (ADR). The stage was cooled to 103~mK. The input and output were connected to a vector network analyzer (VNA) which sent a -40~dB signal to the MKID.  The signal is amplified by a cryogenic  
High Electron Mobility Transistor (HEMT) amplifier.

We simulated the 12 resonators previous to the fabrication to know their resonance frequencies, load quality factors, geometrical capacitance and geometrical inductance. The simulations of each of the 12 resonators are done separately in SONNET. Fig.~\ref{fig:sonnetsimulatio} shows the geometry used for the simulation of one resonator. The input parameters for the 12 resonators were the same. The stackup from bottom to top, was 500 $\mu m$ silicon (Erel~=~11.7) beneath 1000 $\mu m$ air (Erel~=~1). The metallization on top of the silicon layer was TiN (surface resistance~=~0, thickness~=~0, and the kinetic inductance swept from 15~pH/sq to 40~pH/sq in steps of 2~pH/sq). There was a 50~$\Omega$ port on each end of the  transmission line. From the simulation, we obtain the complex transmission amplitude $S_{12}$ (Figs.~\ref{fig:TranssmissionAmplitude} and \ref{fig:ComplexS12}) and by doing a fit to  equation~\eqref{GaoEquation}, extracted the resonance frequency and loaded quality factor $Q_{l}$. Using  $f_{r}$ and $L_{k}$, $L_{g}$ and $C_{g}$ were obtained by a fit to equation~\eqref{Frsimulation}. The resonance frequency span of the 12 resonators ranged from 0.5 GHz to 4.5 GHz for kinetic inductance sweep from 15 pH/sq to 40 pH/sq. The loaded quality factors of the 12 resonators can be divided in two groups: six resonators with low $Q_{l} \approx 6~\times 10^{3} $ and six resonators with high $Q_{l} \approx 1.2~\times 10^{5}$.

\section{Measurements}

\begin{figure}[H]
       
        \centering \includegraphics[width=8.5cm, height=7cm]{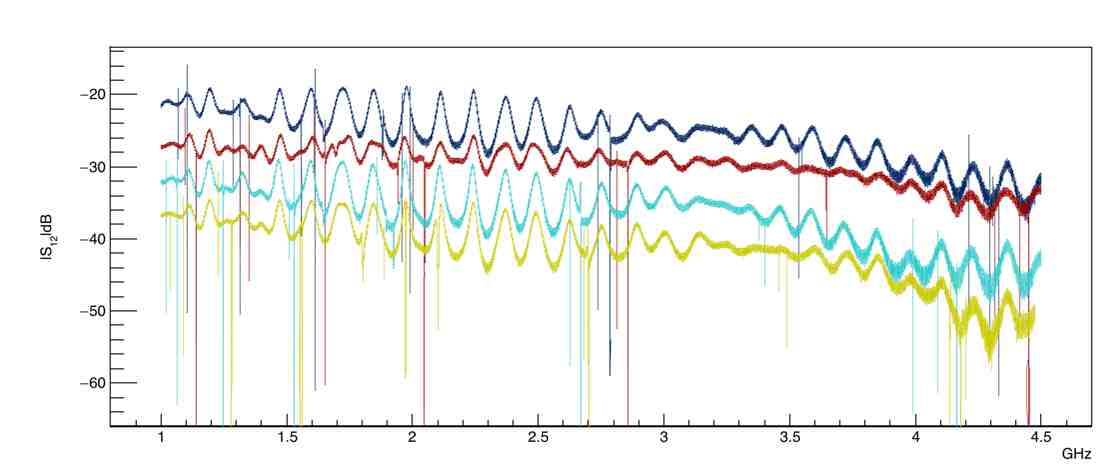}

        \caption{Transmission curves for the four MKIDs at 103~mK and -40~dB. For visualization purposes we add -5~dB, -10~dB and -15~dB shifts to $S_{12}$ for the MKIDs B1, A2 and B2, respectively.}
        \label{fig:span} 
\end{figure}

In Fig.~\ref{fig:span} we show the transmission curves for the four MKIDs obtained at 103~mK and -40~dB. From top to bottom: A1 (blue), B1 (red), A2 (cyan), and B2 (yellow). The resonance frequencies are seen as sharp spikes in this plot. The oscillations and overall shape of the curve are due to the transmission line. The first 10 resonance frequencies are in the range 1--3~GHz and the other two from 4~GHz to 4.5~GHz. The spikes in the range from 3--4~GHz correspond to second modes of the resonance frequencies, as identified in the simulations.

 Analyzing the resonance frequencies, from the four MKIDs A1, A2, B1, B2, we can identify the twelve resonance frequencies in the range  1-4.5 GHz (Fig.~\ref{fig:span}) similar to the simulation, where the MKID A2 contains the lower $f_{r}$ in all its resonators.

\begin{figure}[H]
\includegraphics[width=8.5cm,height=7cm]{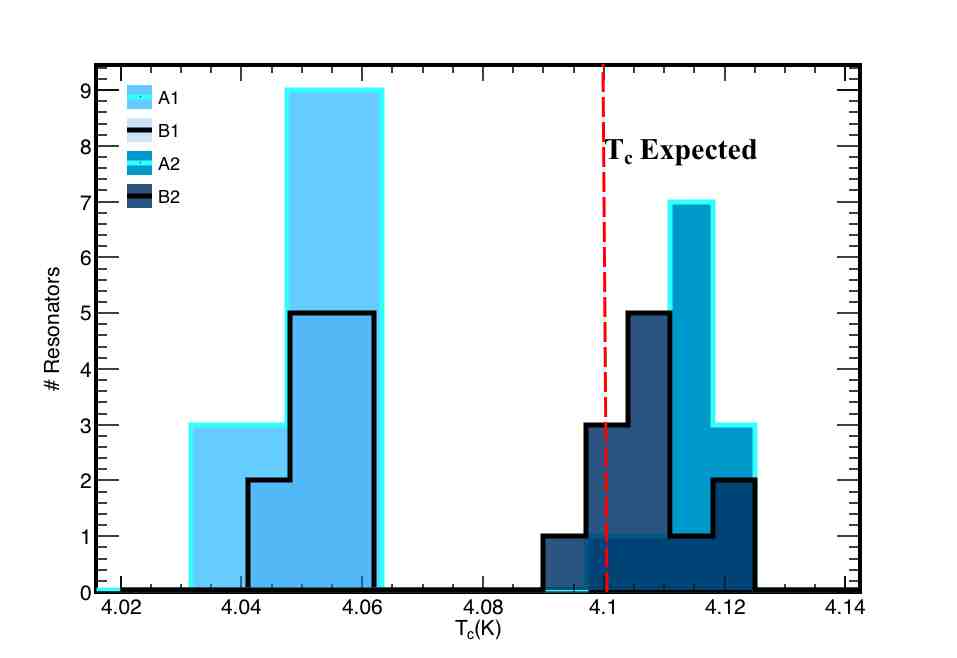}
\caption{
                \label{fig:CriticalTemperature} 
                Critical temperatures distribution  of all 48 resonators obtained using equations~\eqref{Shiftresonance} and \eqref{Mattis}. The dashed red line indicates the expected temperature  $\approx 4.1$K.    
        } 
\end{figure}

To verify the uniformity of ALD we proceed to extract the $T_{c}$  with the method proposed in Appendix~\ref{Critica_temperature}  for all resonators. Fig.~\ref{fig:CriticalTemperature} shows a distribution of critical temperature of all resonators where we can see  four distributions close to the expected critical temperature.  The MKIDs from the first wafer correspond to the left-most distributions and the ones from the second wafer correspond to the right-most distributions. The critical temperature averages are 4.05 K, 4.05 K for B1, A1 and 4.11 K, 4.10 K for A2, B2, respectively.

Calculating the average differences in the critical temperature for the 4 MKIDs (taking as reference the MKID B1) and dividing between the maximum sigma error $\sigma_{T_c}$, we obtain values of $\Delta T_{c}/ \max(\sigma_{1}^T,\sigma_{2}^T)$ of 0.76 and 0.91 for A1-B1 and  A2-B2, respectively. 

 To investigate if the value of $T_{c}$ is constant and independent of the function used to do  the fit, we considered different functions\cite{SuperconductingTiN,GaOthesis} to fit the $\delta f_{r}/f_{r}$ and obtained $T_{c}$. We found that the distribution of the critical temperatures shown in  Fig.~\ref{fig:CriticalTemperature} is the same for all the proposed functions, just varying in the mean $T_{c}$, and the values  of $\Delta T_{c}/ \max(\sigma_{1}^T,\sigma_{2}^T)<1$. This shows that the MKIDs from the same wafer have the same $T_{c}$.

  The percent variations in the critical temperature achieved in sputtering for MKIDs of TiN\cite{TiNCriticalMultilayers,TcSputtering} are of the order of 25\%  and, in our case, the percent variation of A1-B1 and A2-B2 is  0.3\% and 0.5\% respectively. This shows that ALD leads to a better control in the critical temperature, even with MKIDs from different wafers, where the percent variation obtained is  (1.9\%).

\begin{figure}[H]
        \centering
\includegraphics[width=\columnwidth]{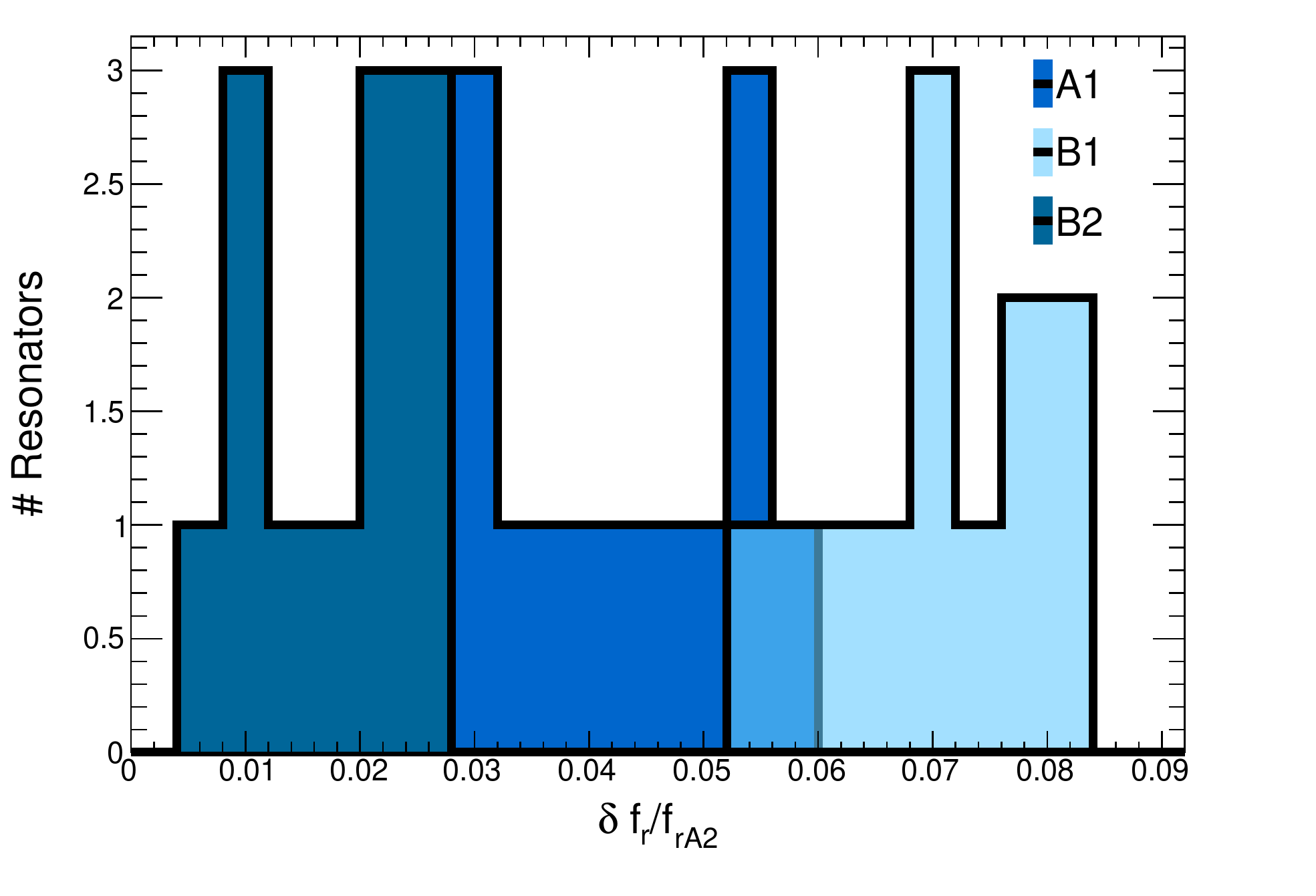}
        \caption{Distribution of $\delta f_{r}/f_{rA2}$.}
        \label{fig:DifferenceResonce} 
\end{figure}

\begin{figure}[H]
       
        \centering \includegraphics[width=8.5cm, height=7cm]{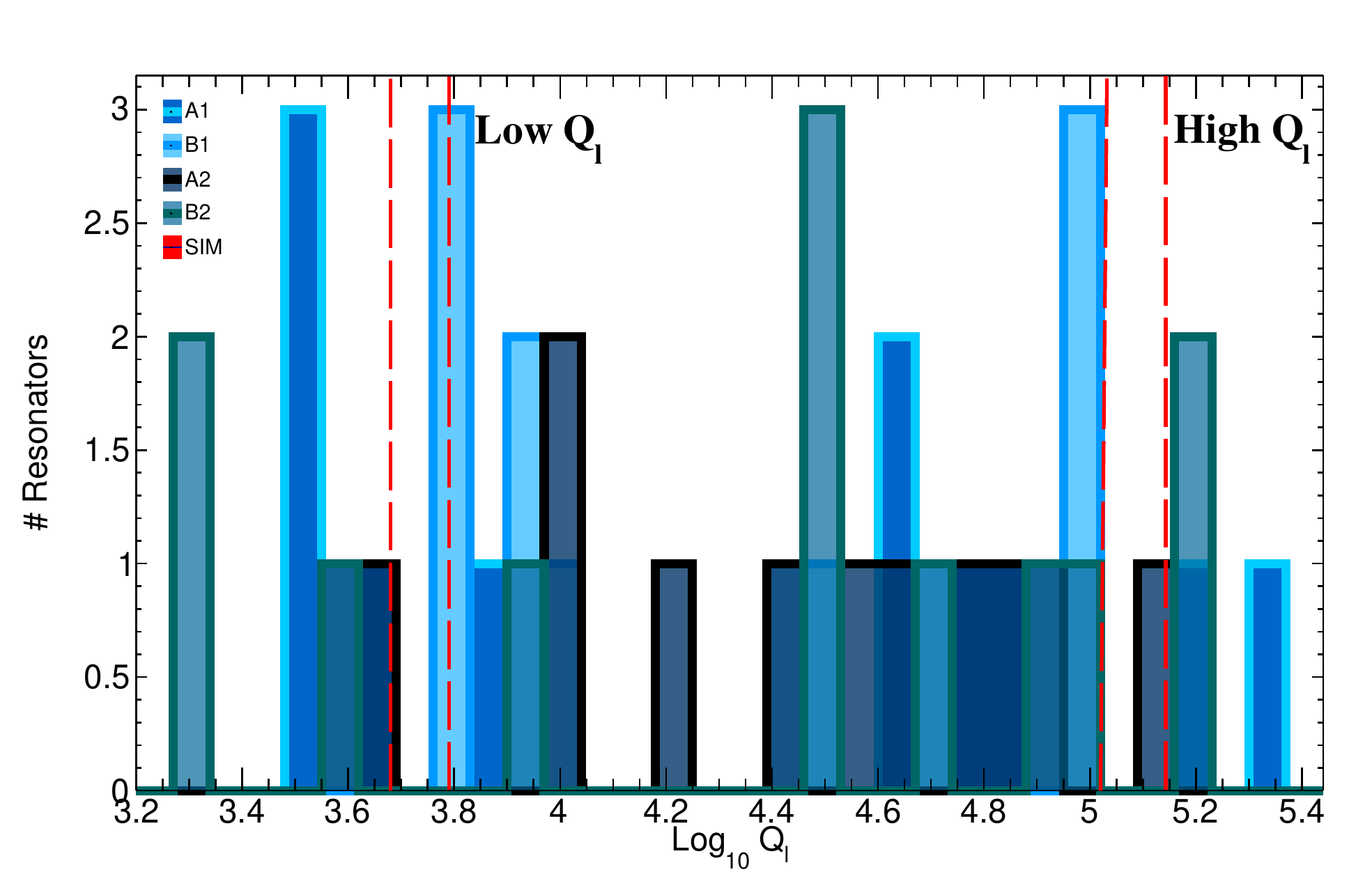}

        \caption{
                \label{fig:Qualitiesfactors} 
                Distribution of  $\log_{10}(Q_{l})$ for the 48 resonators. The dashed red lines correspond to the minimum  and maximum loaded quality factors for the groups with high $Q_{l}$ and low $Q_{l}$.
        }
\end{figure}

Notwithstanding this low percent variation, it is obtained a high variation in the resonance frequency (Fig.~\ref{fig:DifferenceResonce}) and a discrepancy in the loaded quality factor (Fig.~\ref{fig:Qualitiesfactors}).

 The fractional change  of the resonance  frequency $\delta f_{r}/f_{r A2}$ (comparing the resonators {1,..,12} from the MKID A2 with the resonators {1,..,12}  from the MKID A1, B1, B2) is plotted in a histogram (Fig.~\ref{fig:DifferenceResonce}). The fractional differences are rather small (less than 10\%), being less than 2.6\% for the 2 MKIDs from wafer and  less than 3.3\% for those from wafer 2. The maximum change found in $\delta f_{r}/f_{rA2}$ for A1-B1 is $2.6\times 10^{-2}$  and for A2-B2 is $3.3 \times 10^{-2}$. These values are higher in comparison with the value reported of $1.1 \times 10^{-2}$ for identical resonators of TiN built through sputtering\cite{V_D_F_R_0} (the resonators are on the same wafer and same MKID).

To verify if the changes of the resonance frequencies are  a consequence of the variation of the critical temperature, we analyze the local kinetic inductance of all 48 resonators through the methods of Appendix~\ref{Kinetic_Inductance}. In the case of the local kinetic inductance $ L_ {k_ {l}} $ the values are between 18~pH/sq and 24~pH/sq (Fig.~\ref{fig:Lkindividual}), where each resonator has a unique value of $L_ {k_ {l}} $. Calculating  the global kinetic inductance $ L_ {k_ {G}} $ using equation~\eqref{eq:Globlal_kinetic}, the values are 19.5~pH/sq, 20.4~pH/sq, 21.5~pH/sq, 22.13~pH/sq for B1, A1, B2, A2, respectively. Then, we obtained a change in $T_{c}$ of 0.5~K for A1 respect to A2 and 0.1 K for A1,B1 and A2,B2 considering that $ L_{k} \propto 1/T_{c}$. Those values are bigger in comparison with the expected values obtained through the fit to the fractional resonance frequency change, given in table \ref{tab:Tc}. Therefore, the variation in the resonance frequency could be due to other factors such as the variation in the losses of the system and in the over-etch into silicon \cite{V_D_F_R_0}.

 In a resonator the losses of the system are counted by the loaded quality factor $Q_{l} $. From Fig.~\ref{fig:Qualitiesfactors} it is clear that each resonator has a different $Q_{l} $, and in the half of them its value is below the one obtained from the simulation. The fact that the loaded quality factor results lower than the simulated value indicates that there are more losses in the system. These could be by the wire bonding, the connection with the ports and other external agents. In contrast, the values above the designed value could be produced by the CPW line. If we observe the Fig.~\ref{fig:span} the variation of the $S_{12}$~dB  presented in the baseline is produced by a mismatch with the CPW line and its connections. 

In order to explain any possible process involved in the fabrication stages, we performed three simulations, changing, 1) the area of the resonators by $\pm$ 10\% (including the CPW and all the components), 2) the dielectric permeability of the substrate (from 68\% to 171\%)  and 3) the non-homogeneity of the etch (depth from 0 to 150~$\mu m$). In all the simulations the kinetic inductance used was 21~pH/sq and, in 2) and 3) the Erel was taken as 11.7.

The simulation 1) shows a change of  $\sim \pm $9\% in $Q_{l}$ and $ \pm 57$ MHz in $f_{r}$. In 2), $Q_{l}$  varies from 128\% to 76\%  and $f_{r}$ from 233~MHz to -213~MHz. Finally, 3) shows a decrease in $Q_{l}$ of 50\%  for 10~$\mu m$  and 24\% for 150~$\mu m$, and  the resonance frequency  changed from  1.1 GHz to 2.24 GHz. All these simulations represent extreme cases and the changes considered are way above the specifications of all the processes involved in the fabrication. For example, the reported purity of Si is around 99.99\% \cite{SiliconBOOk}, which mean that the permeability can change by at most 0.001\%.

\section{Conclusions}
We have demonstrated the uniformity of Atomic Layer Deposition through measuring the critical temperature of 48 resonators of TiN. Indistinguishability in the critical temperature in resonators from the same process and a difference of 66 mK between different process are obtained. In this way, ALD significantly improves the non-uniformity issues in the critical temperature for the fabrication of the MKIDs and superconducting devices. However, in spite of the uniformity, we found a bigger variation in the resonance frequencies in comparison with sputtering. In the same way, the variations of the loaded quality factor are completely unpredictable in each resonator and complicated to control. As a future work, we will investigate the process that affects both parameters.

{\bf Acknowledgements:} This work was in partial fulfillment of the requirements for the Degree of Master in Physics  at Universidad de Guanajuato. We thank the Pritzker Nanofabrication Facility (PNF) at University of Chicago for access to the facility and  people for their support [including Amy Tang for the dicing]. We appreciate the valuable comments of Dario Rodriguez and Brenda Cervantes. IH and MM acknowledge the hospitality of Fermilab, where this work was done.

\appendix

\section{Fitting $S_{12}$} \label{FittoS12}
In this appendix, we discuss how to obtain the resonance frequency and loaded quality factor from a resonator. Equation~\eqref{GaoEquation} contains  7 parameters and two terms, one associated to the environment and the  other to the properties of the resonator. The environment term accounts for the contributions to the off-resonance frequencies of $S_{12}$. This term depends on a complex constant   $a$ accounting for the gain and phase shift through the system and  $\tau$ describing the cable delay. The environment is analytically modeled as a Fourier series expansion \cite{Analysimethodgoodfunction}, but, in our case, we take only the first term because we are taking data around the minimum. Terms of higher order are not necessary  as they are used to model all the cable in a bigger range.   

To obtain the values of the resonance frequency and loaded quality factor of each resonator we fit the measured transmission amplitude $S_{12}^{\rm data}$ with equation~\eqref{GaoEquation} by minimizing the function

\begin{equation}
\chi^{2}=\sum_{n=0}^{N} \frac{|{S_{12}(f_{n})-S_{12}^{\rm data}(f_{n})}|^{2}}{\sigma_{n}^{2}}\,,
\label{GaoEquationFit}
\end{equation}
where  $\sigma_{n}^{2}$  is the associated error in each point of $S_{12}$ obtained from the network analyzer. 

Equation~\eqref{GaoEquation} depends on the initial values of the seven parameters and $\sigma_{n}^{2}$. To calculate $\sigma_{n}$, we do a new smoothed curve where each point is defined as the mean of the $2n$ nearest neighbors ($n=8$ in our case). The associated error  is computed as the root median square (rms) around the mean. We observe that the data fluctuations lay between the error curves.  

We use different methods\cite{DifferentsMethods} to determine the initial values of the seven parameters. The values of $a$ and $\tau$ are obtained by doing a fit to the environment term $a\exp{(-2\pi j f\tau)}$, in ranges outside the resonance frequency. The initial value of the  resonance frequency is  the point in the complex plane of $S_{12}$ with  maximum  phase velocity. The initial values of the loaded quality factor and  $\phi$  are obtained via the -3~dB method and the phase vs frequency method, respectively. 

\begin{figure}[H]
        \centering \includegraphics[width=\columnwidth]{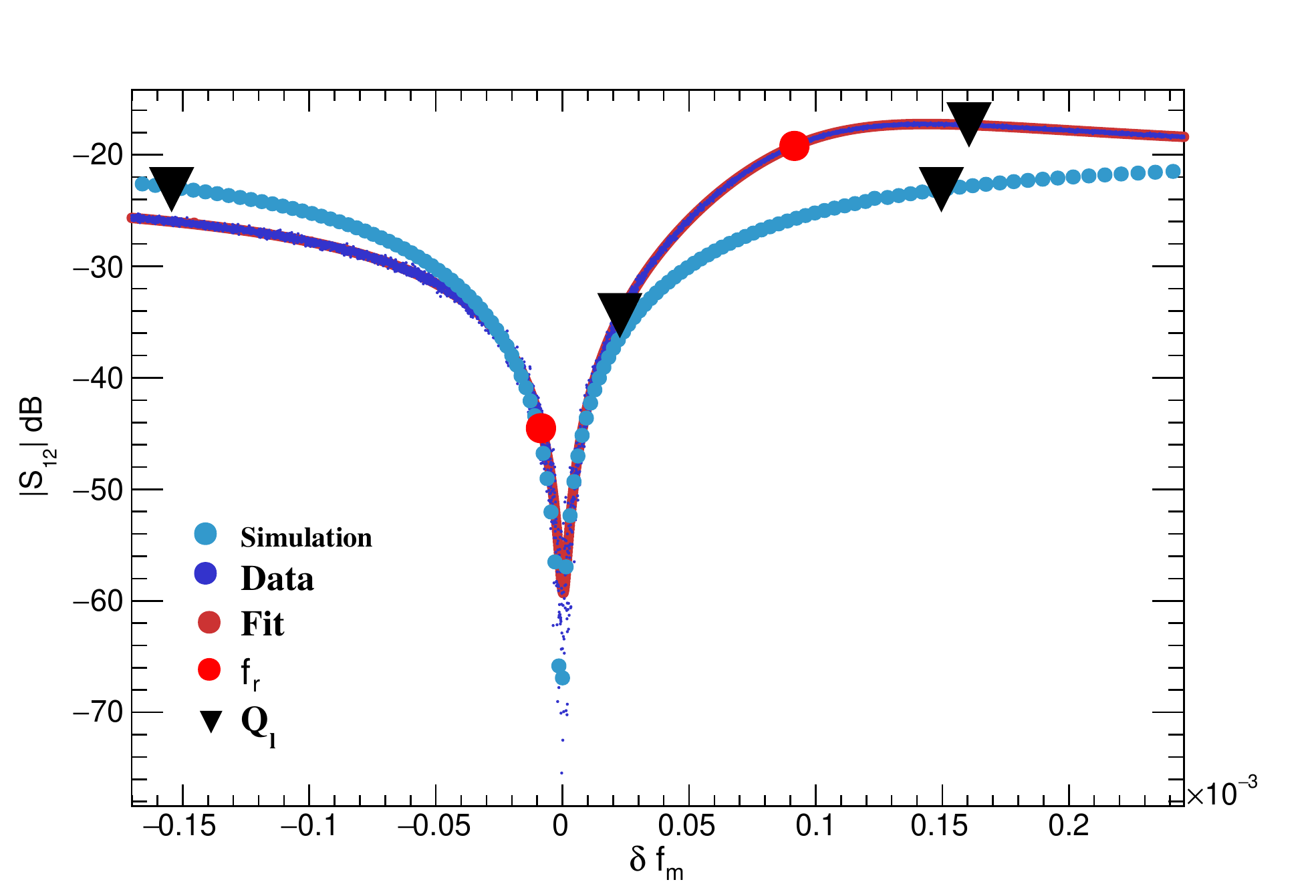}
        \caption{Transmission amplitude of the data, fit and simulation. The black triangles are the coordinates of the bandwidth and the red circles correspond to the resonance frequencies. The data obtained from the simulation are multiplied by  the term $ae^{-2\pi j f \tau}$ obtained from the fit to the experimental data of the resonator. The minimum of each curve of $S_{12}$~dB is moved to 0.}
        \label{fig:TranssmissionAmplitude}
\end{figure}

\begin{figure}[H]
       \includegraphics[width=\columnwidth]{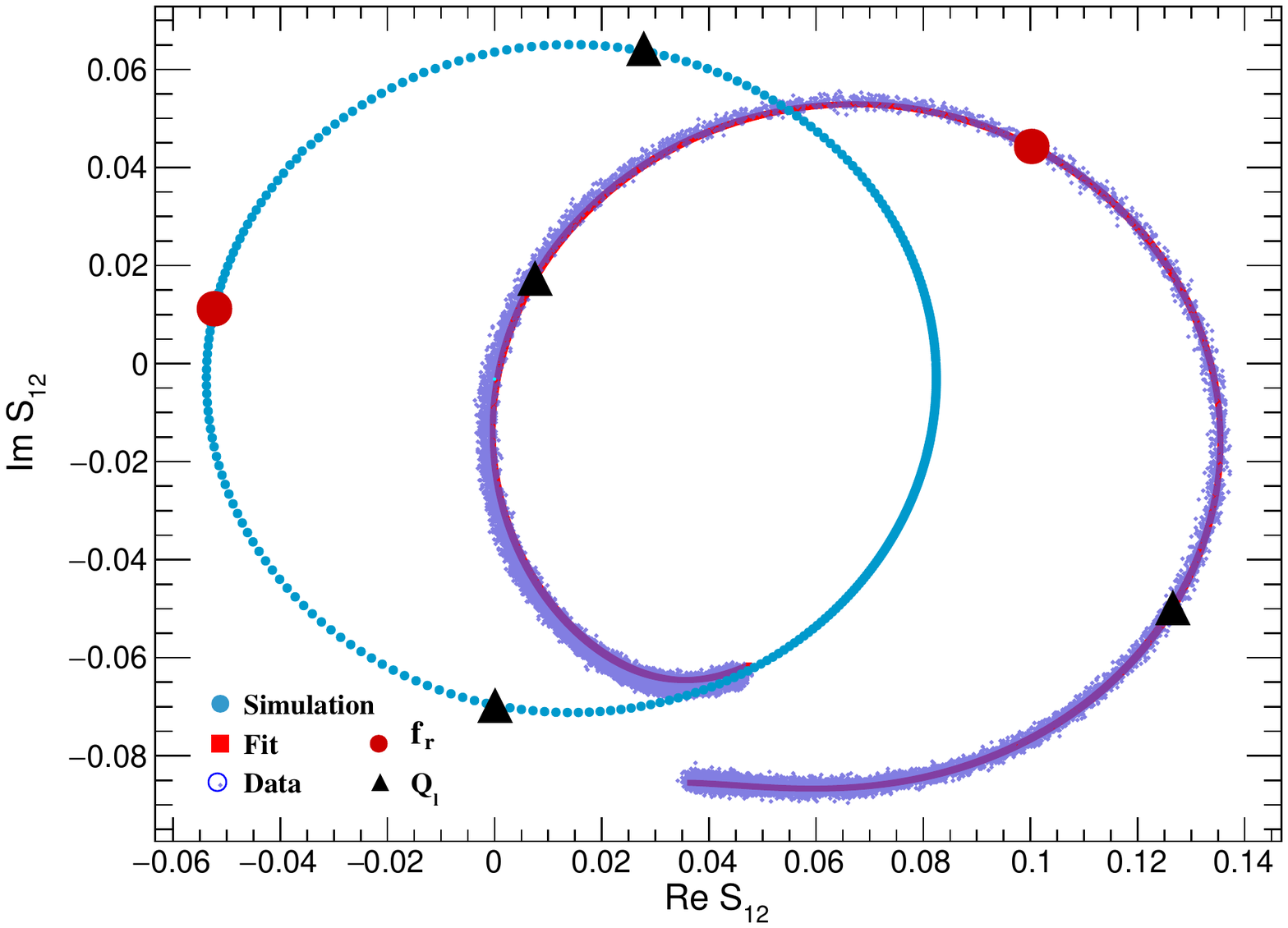}
        \caption{Complex transmission amplitude simulated, measured and fitted. The red circles are the resonance frequencies  and the black triangles correspond to the bandwidth coordinates. The simulated data are multiplied by  the term $ae^{-2\pi j f \tau}$.}
         \label{fig:ComplexS12} 
\end{figure}

 In Fig.~\ref{fig:TranssmissionAmplitude} we observe that the resonance frequency of the data appears above the baseline and the same for  the bandwidth $\Delta f$. Due to our resonators have  an asymmetry in the transmission amplitude  $S_{12}$~dB, the resonance frequency is not in the minimum and the $\Delta f$ is not at -3~dB respect to the baseline. In Fig.~\ref{fig:ComplexS12}, the $f_{r}$ is always in the point with maximum phase velocity and the bandwidth divides the circle by the half. Those properties are observed in all the
 resonators,  independently if they are simulated or experimentally measured. 
 
We use TMINUIT with the initial values of all parameters to obtain the resonance frequency and loaded quality factor for each resonator. Next, we define the error of the fit as
\begin{equation}
\sigma_{error}=|S_{12}^{data}(min(f_{n}))|-|S_{12}(min(f_{n}))|\,.
\label{eq:Errofit}
\end{equation}
In this way, we can establish that two or more resonators are identical if the difference between their resonance frequencies is lower than their maximum $\sigma_{error}$.

\begin{table*}[]
\begin{ruledtabular}
\begin{tabular}{c|cc|cc|cc|cc}
            & \multicolumn{8}{c}{MKIDs} \\
\hline 
& \multicolumn{2}{c|}{A1} & \multicolumn{2}{c|}{B1} & \multicolumn{2}{c|}{A2} & \multicolumn{2}{c}{B2} \\
\hline 
Resonator & $f_{r}^{data}$ (GHz)        & $Q_{l} \times 10^3$        & $f_{r}^{data}$ (GHz)         & $Q_{l}$        & $f_{r}^{data}$ (GHz)        & $Q_{l} \times 10^3$         & $f_{r}^{data}$ (GHz)         & $Q_{l} \times 10^3$         \\
\hline
10                    & 1.0697     & 147.6     & 1.0967     & 74.8      & 1.0207     & 103.0     & 1.0376     & 44.7      \\
1                    & 1.1038     & 3.6       & 1.1408     & 10.1      & 1.0645     & 8.0       & 1.0869     & 31.7      \\
12                    & 1.2868     & 31.4      & 1.3148     & 84.8      & 1.2160     & 54.7      & 1.2275     & 155.8     \\
3                    & 1.3136     & 32.2      & 1.3495     & 4.0       & 1.2487     & 8.0       & 1.2813     & 3.2       \\
7                    & 1.5594     & 165.6     & 1.6073     & 61.0      & 1.5113     & 89.4      & 1.5516     & 46.5      \\
5                    & 1.6116     & 8.4       & 1.6532     & 10.0      & 1.5312     & 10.5      & 1.5620     & 3.5       \\
11                    & 1.9603     & 104.3     & 2.0037     & 35.0      & 1.8603     & 95.6      & 1.8872     & 232.6     \\
2                    & 1.9890     & 1.9       & 2.0482     & 16.6      & 1.9212     & 6.5       & 1.9728     & 7.0       \\
4                    & 2.7363     & 54.4      & 2.8119     & 28.0      & 2.6269     & 26.7      & 2.6826     & 70.8      \\
8                   & 2.7868     & 2.0       & 2.8559     & 4.6       & 2.6691     & 6.5       & 2.6996     & 3.5       \\
9                   & 4.2118     & 80.8      & 4.3164     & 125.7     & 4.0898     & 29.1      & 4.1386     & 55.4      \\
6                   & 4.2970     & 31.8      & 4.4136     & 42.0      & 4.1653     & 6.3       & 4.1821     & 4.6      
\end{tabular}
\end{ruledtabular}
\caption{\label{tab:resonace_frequencies} Resonance frequencies and loaded quality factors for the 12 resonators of the 4 MKIDs considered in this work.  }
\end{table*}
\section{Critical Temperature }\label{Critica_temperature}
To extract the critical temperature through the fractional change of the resonance frequency we  perform  a chi-squared fit

\begin{equation}
\chi^{2}=\sum\limits_{i=1}^n    \frac{\left(\frac{\delta f_{r}(T_{i})}{f_{r}}- \frac{\alpha \delta \sigma_2 (T_{i},T_{c},\omega_{i})}{\sigma_2(0)} \right)^{2}}{\sigma_{\delta f_{r}/f_{r}}^{2}}\,,
\end{equation}

  where the error  $\sigma_{\delta f_{r}/f_{r}}$ is  the median absolute deviation (MAD) and is global. The data set of the MAD is obtained from the error propagation of $\delta f_{r}(T_{i})/f_{r}(0)$.

\begin{figure}[H]
        \centering \includegraphics[width=\columnwidth]{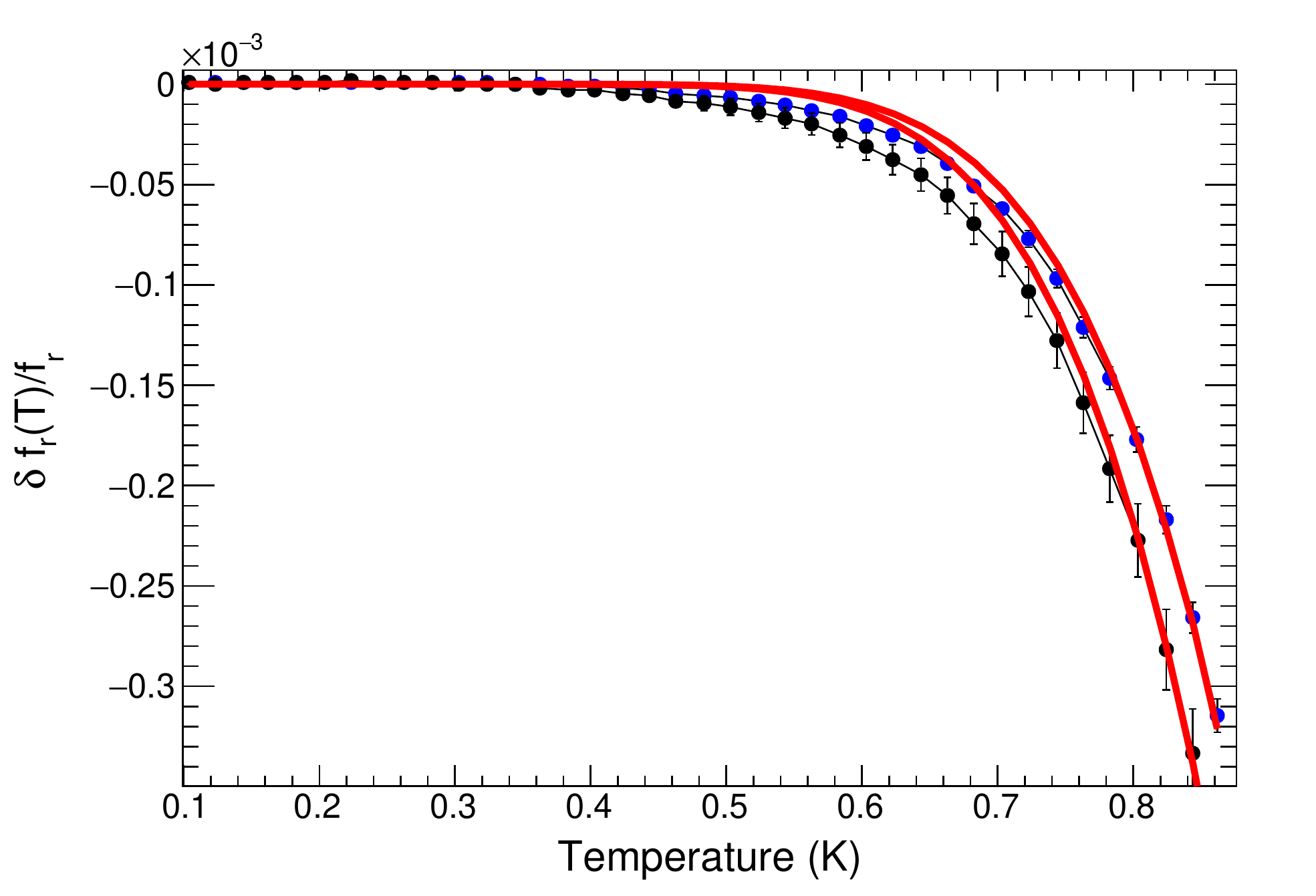}
        \caption{Fractional  frequency change $\delta f_{r}/f_{r}$ as a function of the temperature for the resonator 2 from the MKID A1 (black) and MKID A2 (blue). The red lines are the fit to  equation \eqref{Shiftresonance} using the complex conductivity in equation \eqref{Mattis}.}
        \label{fig:CriticaltemperatureFit}
\end{figure}

\begin{table}[]

\begin{ruledtabular}
\begin{tabular}{c|cccc}
& \multicolumn{4}{c}{$T_{c}$ (K)}                  \\
\hline 
Resonator   & A1 & B1 & A2 & B2 \\
\hline 
10                    & 4.05        & 4.06        & 4.12        & 4.11        \\
1                    & 4.05        & 4.06        & 4.12        & 4.11        \\
12                    & 4.05        & 4.06        & 4.12        & 4.12        \\
3                    & 4.05        & 4.06        & 4.12        & 4.10        \\
7                    & 4.05        & 4.05        & 4.12        & 4.11        \\
5                    & 4.06        & 4.06        & 4.11        & 4.10        \\
11                    & 4.05        & 4.05        & 4.12        & 4.12        \\
2                    & 4.05        & 4.05        & 4.12        & 4.11        \\
4                    & 4.05        & 4.05        & 4.11        & 4.10        \\
8                   & 4.05        & 4.05        & 4.11        & 4.10        \\
9                   & 4.04        & 4.05        & 4.10        & 4.10        \\
6                   & 4.05        & 4.05        & 4.11        & 4.09       
\end{tabular}

\end{ruledtabular}
\caption{\label{tab:Tc}Critical temperature of the 48 resonators from the 4 MKIDs obtained with equations \eqref{Shiftresonance} and \eqref{Mattis}.  }
\end{table}

\section{Kinetic Inductance }\label{Kinetic_Inductance}
From equation~\eqref{Frsimulation}, we can derive the kinetic inductance of a resonator $L_{k}$ using the values of $f_{r}$, $L_{g}$ and $C_{g}$. Depending on the design of the resonators,  in some cases it is possible to determine the values of $L_{g}$ and $C_{g}$  by  analytical methods \cite{Anlatyc_Lg}. However, in our case, it is complicated due to the complexity of  the design.
 
 Running a simulation with different values of $L_{k}$, from 15~pH/sq to 40~pH/sq in steps of 2~pH/sq, we obtained $f_{r}$ as a function of  $L_{k}$. Then, we did a fit to equation~\eqref{Frsimulation} and the values of the geometrical capacitance $C_{g}$ and geometrical inductance $L_{g}$ of each resonator were extracted. From the simulation we have identified the 12 resonators and the order in which they appear in the frequency space, which is independent of the value of $L_{k}$. This allows us to identify to which resonator corresponds each resonance frequency when we analyze the spectrum of the MKID (see Fig.~\ref{fig:span}).

\begin{figure}[H]
       
        \centering \includegraphics[width=\columnwidth]{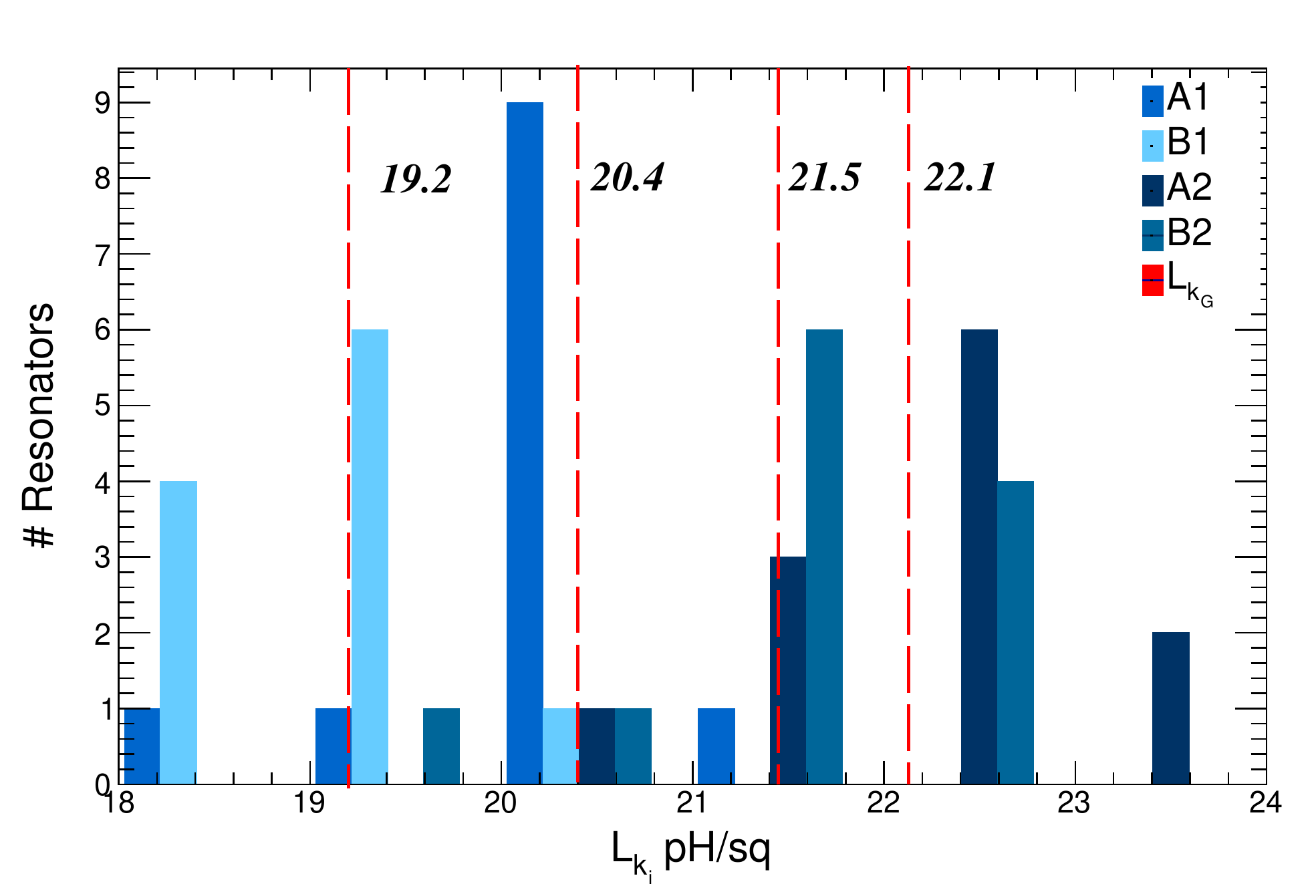}

        \caption{
                \label{fig:Lkindividual} Distribution of the local kinetic inductance extracted from the equation~\eqref{Frsimulation} for the  48 resonators. The dashed red lines are the global kinetic inductance of each MKID.
        }
\end{figure}

 Fig.~\ref{fig:Lkindividual} shows a histogram of the local kinetic inductance $L_{k_{l}}$ of the 48 resonators derived from the measured value of $f_{r}$ and the simulated values of $L_{g}$ and $C_{g}$. We can obtain a unique value of the kinetic inductance of all resonators, referred to as  global kinetic inductance  $L_{k_{G}}$, through finding the value of  $L_{k}$ that minimizes the next function
\begin{equation}
\alpha(L_{k})=\sum\limits_{i=1}^n \frac{(f_{r_{i}}^{data}-f_{r_{i}}(L_{k}))^{2}}{\sigma_{error}^{2}}\,.
\label{eq:Globlal_kinetic}
\end{equation}
In the equation~\eqref{eq:Globlal_kinetic} the sum runs over all 12 resonance frequencies and the error  $\sigma_{error}^{2}$ is calculated with equation~\eqref{eq:Errofit}. 

Fig.~\ref{fig:Lkindividual} shows the values of  the global kinetic inductance derived  with the previous methods.

\subsection{\label{sec:citeref}Citations and References}

\bibliography{MKIDbiblio}

\end{document}